\documentclass[conference,compsoc]{IEEEtran}
\usepackage[section]{placeins}
\usepackage{float}
\usepackage{graphicx}
\graphicspath{ {./figures/} }

\ifCLASSOPTIONcompsoc
  \usepackage[nocompress]{cite}
\else
  \usepackage{cite}
\fi
\ifCLASSINFOpdf
\else
\fi
\begin{document}
%
\title{Benchmarking GPUs on SVBRDF Extractor Model}

\author{\IEEEauthorblockN{Narayan Kandel}
\IEEEauthorblockA{School of Computing\\
Clemson University\\
Email: nKandel@g.clemson.edu}
\and
\IEEEauthorblockN{Melanie Lambert}
\IEEEauthorblockA{School of Computing\\
Clemson University\\
Email: melambe@g.clemson.edu
}}


%


\maketitle

\begin{abstract}
With the maturity of deep learning, its use is emerging in every field. Also, as different types of GPUs are becoming more available in the markets, it creates a difficult decision for users.  How can users select GPUs to achieve optimal performance for a specific task? Analysis of GPU architecture is well studied, but existing works that benchmark GPUs do not study tasks for networks with significantly larger input. In this work, we tried to differentiate the performance of different GPUs on neural network models that operate on bigger input images (256x256).
\end{abstract}

%
\IEEEpeerreviewmaketitle

\section{Introduction}
The growth of research in deep learning is causing it to be relevant in numerous different fields. Deep learning requires computationally-intensive training and utilizes high-performance computing (HPC) to handle large amounts of data. Graphics processing units (GPUs) are perfect for HPC because of their abundance of cores and potential for memory. Leading manufacturers in the GPU industry include NVIDIA and AMD, whose GPUs are used in top-performing supercomputers. As of 2021, in the Top500  list \cite{top500}, the world's fastest supercomputer is Fukagu, with 7,630,848 using 27,000 NVIDIA Tesla GPUs linked to 9,000 IBM Power9 CPUs. NVIDIA GPUs are used in all the top three supercomputers. The fifth highest is Perlmutter and uses AMD EPYC based nodes, and 1536 NVIDIA A100 accelerated nodes. Fukagu achieves 442 Pflop/s, and Perlmutter achieves 64.6. We will be focusing on NVIDIA GPUs in our experiments for our purposes and because of availability. 

As varying GPU architecture becomes more available, it can be a challenge to get optimal performance of GPU. In order to approach the problem of getting optimal value, we present a benchmarking report. We compare the performance of an extensive network with sizable input/output features on a CPU and different GPUs. 

In this work, we do a comprehensive analysis regarding the performance of a neural network with significantly large input/output features. This gives a better understanding of resource usage and presents techniques for optimal performance in the visual computing field. Additionally, this comparative report on power usage, memory usage, and time-required comparison between GPU machines provides valuable information when making decisions regarding hardware selection and optimum utilization.

We organize this paper according to the following. The literature review is discussed in section 2, and in section 3, we outline our methodology. In section 4, we discuss and analyze results, comparing different features of each GPU and their impact on performance. Finally, we draw our conclusion in section 5.

\section{Literature Review}
Analysis of GPU architecture is a well-studied problem, and some existing works analyze the architectural differences between GPUs and their impact on performance (Zhang et al.\cite{zhang}, Wang et al. \cite{wang}). These approaches generally test on tasks such as matrix multiplication and clustering jobs. However, there is no known work benchmarking for networks with bigger images such as 256x256 in our case. 

Zhang et al. \cite{zhang} provide a comparison of older AMD and NVIDIA GPU architectures. Specifically, they provide an in-depth study to distinguish specific differences in NVIDIA's Fermi and ATI's Cypress architecture and how they impact performance. They also compare energy efficiencies, an area of great interest in high-performance computing. They find that Cypress is more energy-efficient if the program is a perfect fit for the processors; otherwise, the Fermi is the more energy-efficient choice. Additionally, their conclusion shows that both architectures work better for different tasks, as they have differing amounts of cores and memory subsystems. While this study is on older architectures, it provides an in-depth analysis and comparison of two GPU architectures.

A more recent work Wang et al. \cite{wang} performs an extensive benchmarking analysis of a deep neural network on TPU, GPU, and CPU architectures. They identify a memory bandwidth bottleneck of the TPU architecture and conclude that improving memory access for memory-bound operations is necessary for performance improvement. They also compare the hardware and software of the varying platforms. 

While existing research studies different architectures, their efficiency, and their influence on performance, there is no known work benchmarking for networks with bigger images such as 256x256 in our case. 

\section{Methodology}

\begin{figure}[!tbp]
\centering
\includegraphics[scale=.33]{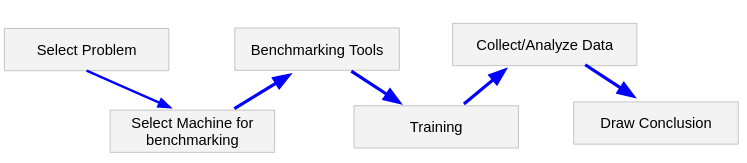}
\caption{Block Diagram}
\label{fig:method_block_diagram}
\end{figure}

In order to do GPU benchmarking, we followed six steps. Those are: 
\begin{enumerate}
\item Selecting Problem
\item Selecting machine for benchmarking
\item Selecting Benchmarking Tools
\item Performing Neural Network Training
\item Collecting/Analyzing data
\item Drawing conclusion
\end{enumerate}

\begin{figure}[!tbp]
\centering
\includegraphics[scale=.45]{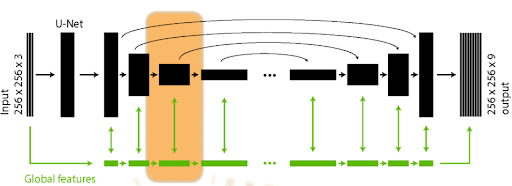}
\caption{Model Structure}
\label{fig:model_structure}
\end{figure}

The first step is selecting the problem. In this work, our objective is benchmarking training performance of the Spatially Varying Bidirectional Reflectance Distribution Function (SVBRDF) Extractor Model. This model is proposed by V. Deschaintre et al \cite{desch}. The basic block diagram of the model is shown in figure~\ref{fig:model_structure}. 

\begin{figure}[!tbp]
  \centering
  \begin{minipage}[b]{0.2\textwidth}
    \includegraphics[width=\textwidth]{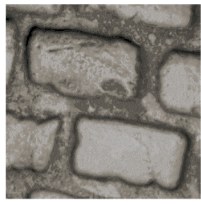}
    \caption{Input Image\\}
    \label{fig:input_image}
  \end{minipage}
  \hfill
  \begin{minipage}[b]{0.2\textwidth}
    \includegraphics[width=\textwidth]{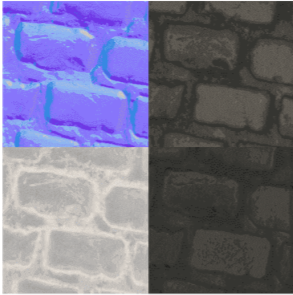}
    \caption{Output as an Image}
    \label{fig:method_output_as_image}
  \end{minipage}
\end{figure}

This model has a total of 80,505,488 parameters and takes 256x256 RGB images as an input. One of the input samples is shown in figure~\ref{fig:input_image}. The output of the network is of size 256x256x9. The result is partitioned into four parts. They are Surface Normal, Roughness, Specular Albedo, and Diffuse Albedo. One sample of output is shown in figure~\ref{fig:method_output_as_image}. 

We used a total of 75 GBs of training data. It consists of approximately 200k images. 

The second step is selecting a machine for benchmarking. Based on the availability of the machine on Palmetto Cluster and usability of the machine, we had selected a single-core GPU machine of type K40, P100, V100, and A100, and a single-node 24 core CPU machine. The basic architecture difference between selected GPU is shown in the table~\ref{table:1}.

\begin{table}[h!]
\begin{center}
\begin{tabular}{ | p{2cm} ||p{0.7cm}|p{1.2cm}|p{1.2cm}|p{1.1cm}|}
\hline
& K40 & P100 & V100 & A100 \\ 
\hline
\hline
Architecture & Kepler & Pascal & Volta & Ampere \\  
\hline
Cuda Core & 2880 & 3584 & 5120 & 6912 \\
\hline
ROP & 48 & 96 & 128 & - \\
\hline
Base Clock (MHz)&745&1189(1324)&1246(1380)&765(1410)\\
\hline
Power (watt)& 245 & 250 & 250 & 400 \\
\hline
Memory Size (MB)& 12288 & 12288 & 16384 & 40960 \\
\hline
Memory Clock (MHz) & 1502 & 715 & 879&1215\\
\hline
Memory Bandwidth(GB/s) & 288.38 & 549.12 &900.1&1555\\
\hline
Price (\$) & 500 & 5900& 6900& 10000 \\
\hline
Release & 2013 & 2016 & 2017 & 2020\\
\hline
\end{tabular}
\end{center}
\caption{GPU Internal Comparison. (\cite{gpucomp-k40-v100}, \cite{gpucomp-p100-v100})}
\label{table:1}
\end{table}

For the third step, we selected benchmarking tools. We utilized Weight and Bias APIs \cite{wandb} to record performance-related data during our training. Our selection is based on two reasons; firstly, the API has a simple configuration, and secondly, all data can be easily visualized using a provided web platform.

The fourth step is training. We train a neural network model for one epoch on each machine. During the training process, we had collected GPU utilization data, GPU memory data, power usage data, total completion data, image processed per second, etc, and store those data on a weight and bias server. 

In the fifth and sixth steps, we had analyzed those data and used them to do benchmarking. We have presented some of the analysis in the result section.

We had used the Tensorflow framework to build a neural network and a palmetto cluster to train the model. During the neural network training process, we had used a learning rate of 0.00002 and 84 test images for validating the model.

\section{Evaluation and Results}
In order to do a comparison between different configurations, we ran a number of experiments and then recorded performance-related data. We break our experiment into four parts. In the first part, we compared  the performance of K40 GPUs with two different batch sizes. In the second part, we trained the network on different GPUs with batch size 5 and compare their performance. In the third part, we repeated the analysis done on part two with batch size 25. And in the fourth part, we trained the network on a CPU machine and compared its performance with GPU (K40) machine. The network's output is evaluated by comparing it with the output authors has provided.

\begin{figure}[h!]
\includegraphics[scale=.13]{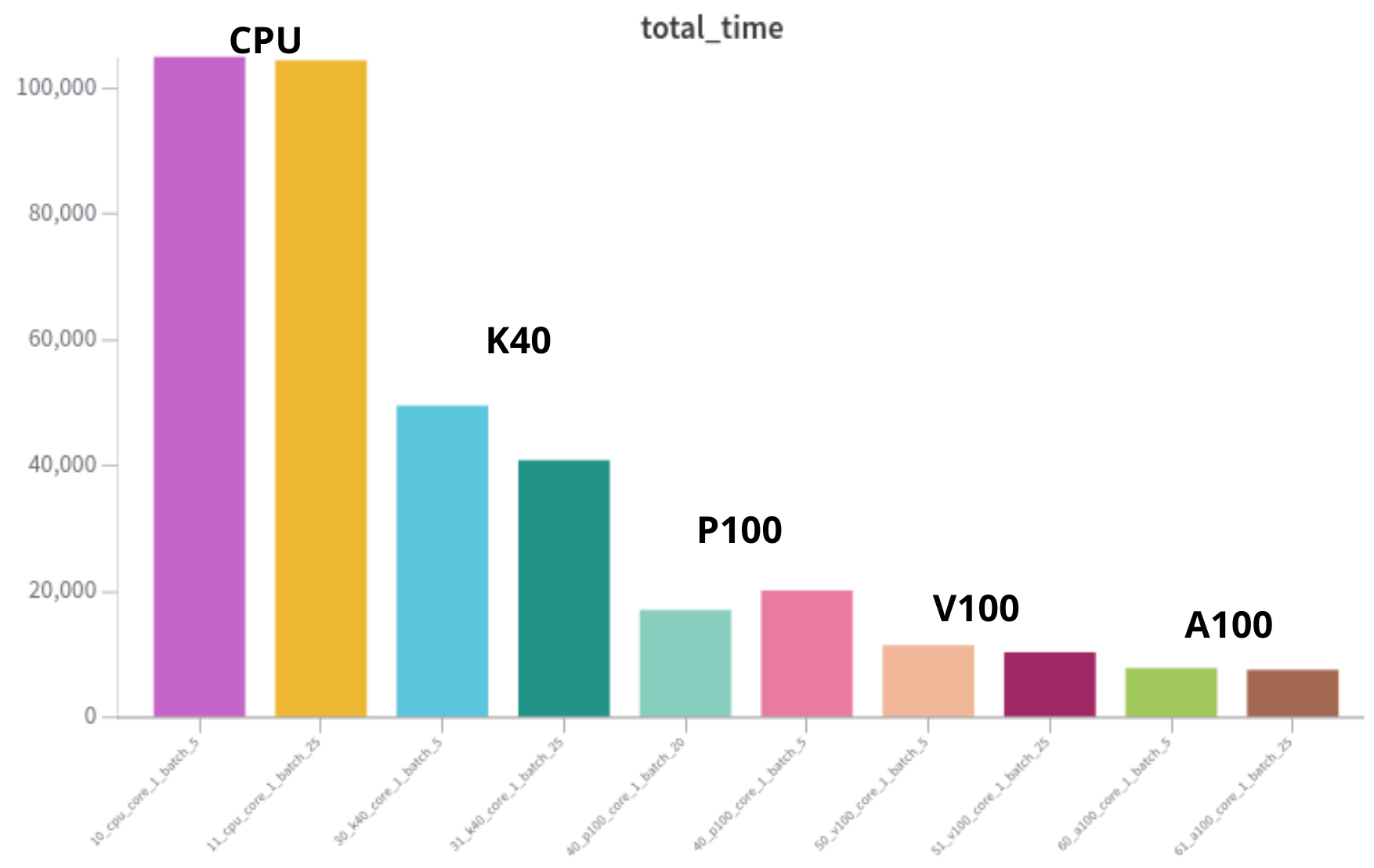}
\caption{Execution Time comparison}
\label{fig:19_time_comparison}
\end{figure}

Figure~\ref{fig:19_time_comparison} shows the total time taken to train a neural network for one epoch. In the figure, the longest time corresponds to the CPU machine with batch size 5 and the shortest time corresponds to the A100 GPU machine with batch size 25. The time trend seen in the figure~\ref{fig:19_time_comparison} matches the computing power of the machine. Among different batch sizes on the same machine, it is seen that training with a higher batch size would reduce the total time requirement by a small margin.    
\subsection{Using different batch sizes on K40 GPU}
\begin{figure}[!tbp]
  \centering
  \begin{minipage}[b]{0.24\textwidth}
    \includegraphics[width=\textwidth]{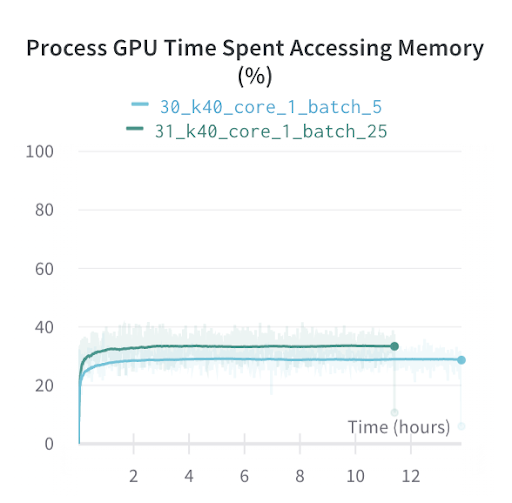}
    \caption{Time spend accessing memory on K40}.
    \label{fig:19_time}
  \end{minipage}
  \hfill
  \begin{minipage}[b]{0.22\textwidth}
    \includegraphics[width=\textwidth]{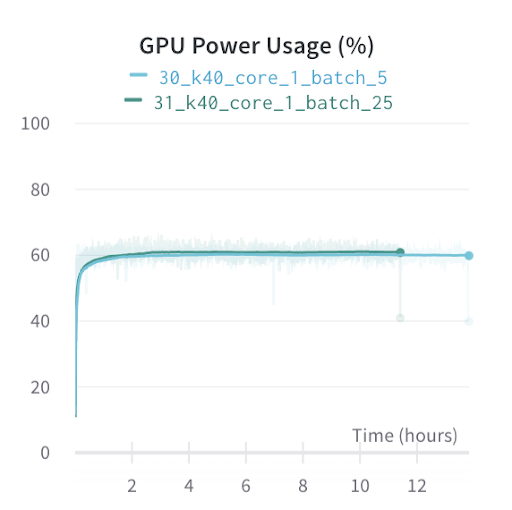}
    \caption{Power Usage on K40}
    \label{fig:19_power}
  \end{minipage}
\end{figure}

\begin{figure}[h!]
\includegraphics[scale=.45]{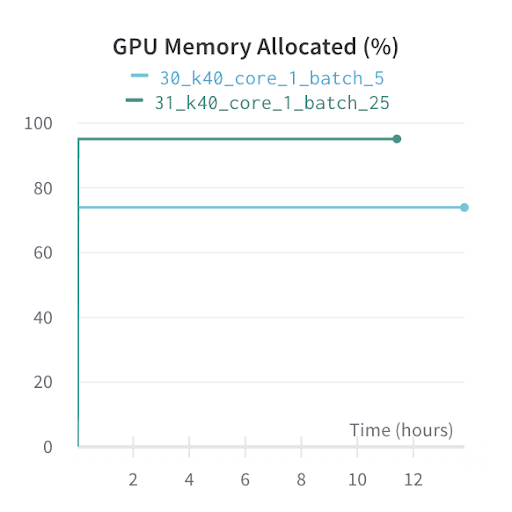}
 \caption{GPU Memory allocation between different batches on K40}
 \label{fig:19_gpu_memory}
 \end{figure}

In the first step, We train networks with batch sizes 5 and 25 on K40 GPU. The GPU memory allocation, process GPU time spent accessing memory, and change in GPU power usage are shown in figure~\ref{fig:19_gpu_memory}, figure~,\ref{fig:19_time} and figure~\ref{fig:19_power}. From the figure, it is observed that processing time spent accessing memory increased with increasing batch sizes though GPU power usage remains constant. In addition, total GPU memory allocation increased with an increase in batch size. In figure~\ref{fig:19_gpu_memory}, we can see that GPU memory allocation reached near 100\% with batch size 25. So, based on the three plots, we can say that we are limited to utilizing full GPU power by GPU memory. 

\subsection{Performance Comparison between GPUs on batch size 5}
\begin{figure}[!tbp]
  \centering
  \begin{minipage}[b]{0.24\textwidth}
    \includegraphics[width=\textwidth]{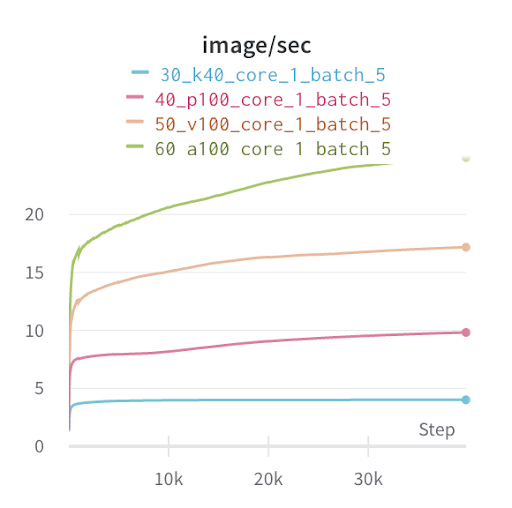}
    \caption{Images process per sec (Batch 5)}
    \label{fig:20_image_rate}
  \end{minipage}
  \hfill
  \begin{minipage}[b]{0.22\textwidth}
    \includegraphics[width=\textwidth]{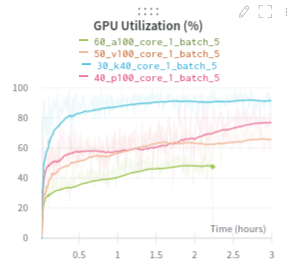}
    \caption{GPU Utilization (Batch 5)}
    \label{fig:20_gpu_utilization}
  \end{minipage}
\end{figure}

\begin{figure}[!tbp]
  \centering
  \begin{minipage}[b]{0.22\textwidth}
    \includegraphics[width=\textwidth]{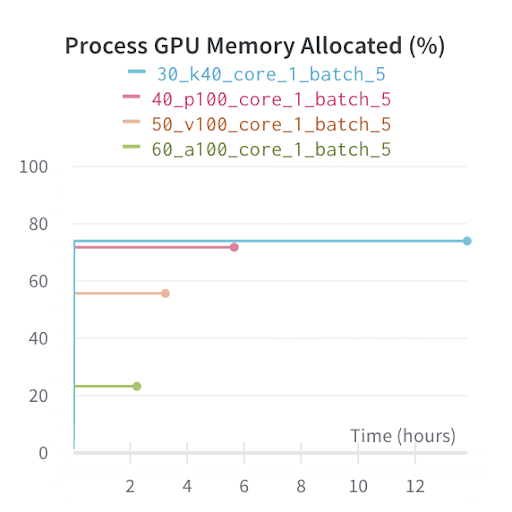}
    \caption{GPU Memory Allocation (Batch 5)}
    \label{fig:20_gpu_memory}
  \end{minipage}
  \hfill
  \begin{minipage}[b]{0.24\textwidth}
    \includegraphics[width=\textwidth]{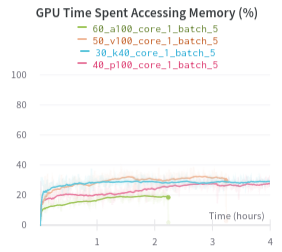}
    \caption{Memory Access Time (Batch 5)}
    \label{fig:20_memory_time}
  \end{minipage}
\end{figure}

In the second part, we ran training across different GPU machines with batch size 5 for one epoch. The number of images processed per second on different GPUs is shown in figure~\ref{fig:20_image_rate}. It is observed that A100 is processing the highest number of images and K40 is processing the lowest number of images even though its GPU utilization isn't near 100\%.  The performance comparison between different GPUs on batch size 5 is shown in figure~\ref{fig:20_image_rate}, figure~\ref{fig:20_gpu_utilization}, figure~\ref{fig:20_gpu_memory} and figure~\ref{fig:20_memory_time}. From these comparison plots, it is observed that A100 is performing better than others as expected and K40 is performing worst among the four GPUs.

\subsection{Performance Comparison between GPUs on batch size 25}
\begin{figure}[!tbp]
  \centering
  \begin{minipage}[b]{0.22\textwidth}
    \includegraphics[width=\textwidth]{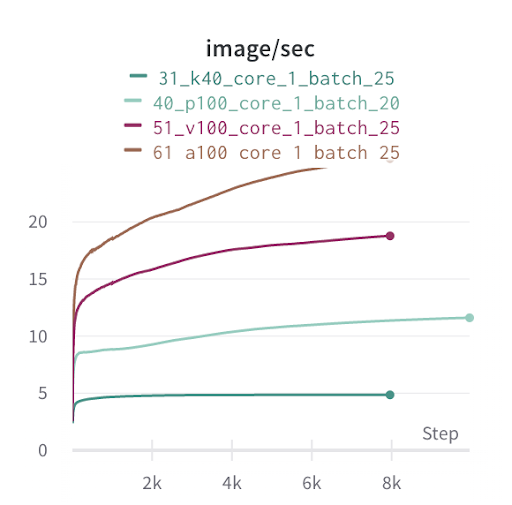}
    \caption{Images process per sec (Batch 25)}
    \label{fig:22_image_rate}
  \end{minipage}
  \hfill
  \begin{minipage}[b]{0.24\textwidth}
    \includegraphics[width=\textwidth]{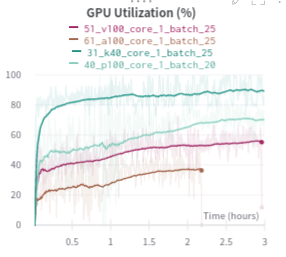}
    \caption{GPU Utilization (Batch 25)}
    \label{fig:22_gpu_utilization}
  \end{minipage}
\end{figure}

In the third part, we trained the network with batch size 25 for one epoch. The pattern on the number of images processed per sec as shown in figure~\ref{fig:22_image_rate} is similar to the pattern on batch size 5. The machine with A100 GPU has the highest image rate and the machine with K40 GPU has the lowest image rate among the four GPUs machine. The Process memory allocation increased with batch size and it reached near 100\% for batch size 25. The GPU utilization plot shown in figure~\ref{fig:22_gpu_utilization} shows that GPUs aren't fully utilized and there is a possibility to get higher performance with an increase in data size. But, as GPU memory already reached near 100\%, we are unable to increase the batch size and hence it acts as one of the walls of the system. 

\begin{figure}[!tbp]
  \centering
  \begin{minipage}[b]{0.22\textwidth}
    \includegraphics[width=\textwidth]{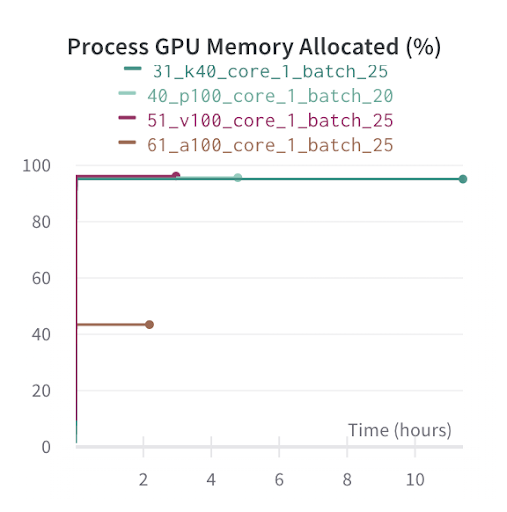}
    \caption{GPU Memory Allocation (Batch 25)}
    \label{fig:22_gpu_memory}
  \end{minipage}
  \hfill
  \begin{minipage}[b]{0.24\textwidth}
    \includegraphics[width=\textwidth]{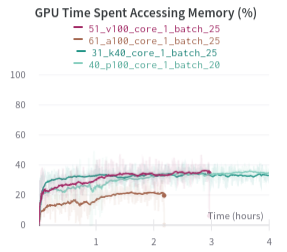}
    \caption{Memory Access Time (Batch 25)}
    \label{fig:22_memory_time}
  \end{minipage}
\end{figure}

Figure~\ref{fig:22_memory_time} shows the Memory access time pattern among four GPUs. As expected, The machine with A100 GPU has lower memory access time and the machine with K40 GPU has higher memory access time.

As an additional observation, Figure~\ref{fig:22_image_rate} shows that A100 GPU and V100 GPU have a much higher image per second rate than K40 GPU. This is expected because A100 and V100 have considerably more render output units (ROPs) than K40. A100 has 160 ROPs, and V100 has 128, whereas K40 only has 49 ROPs. This allows A100 and V100 to draw more images per second, increasing performance. As shown in Figure~\ref{fig:19_time_comparison}, K40 also takes longer than other GPUs because it has the least amount of cores out of all the GPUs.

\subsection{Performance comparison between CPU \& GPU}
\begin{figure}[!tbp]
  \centering
  \begin{minipage}[b]{0.22\textwidth}
    \includegraphics[width=\textwidth]{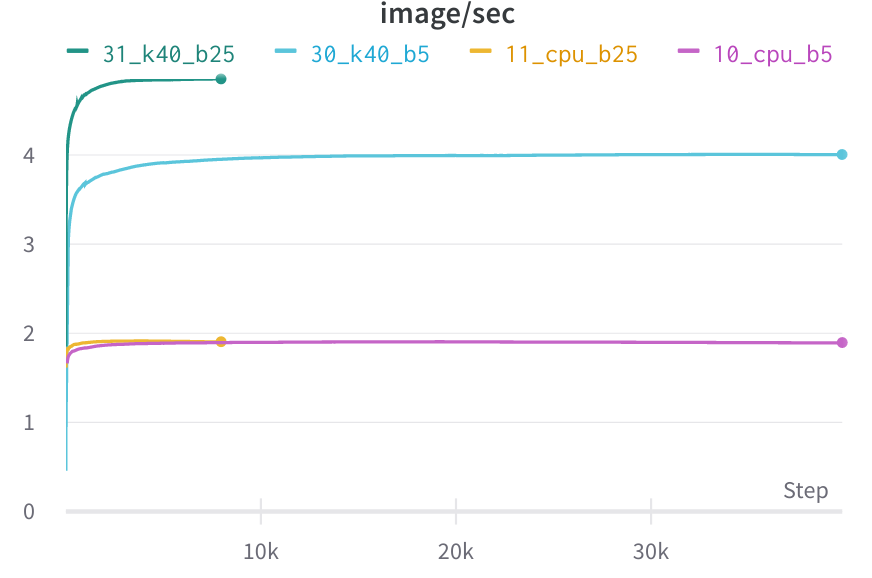}
    \caption{Images process per sec (CPU vs K40)}
    \label{fig:24_image_rate}
  \end{minipage}
  \hfill
  \begin{minipage}[b]{0.23\textwidth}
    \includegraphics[width=\textwidth]{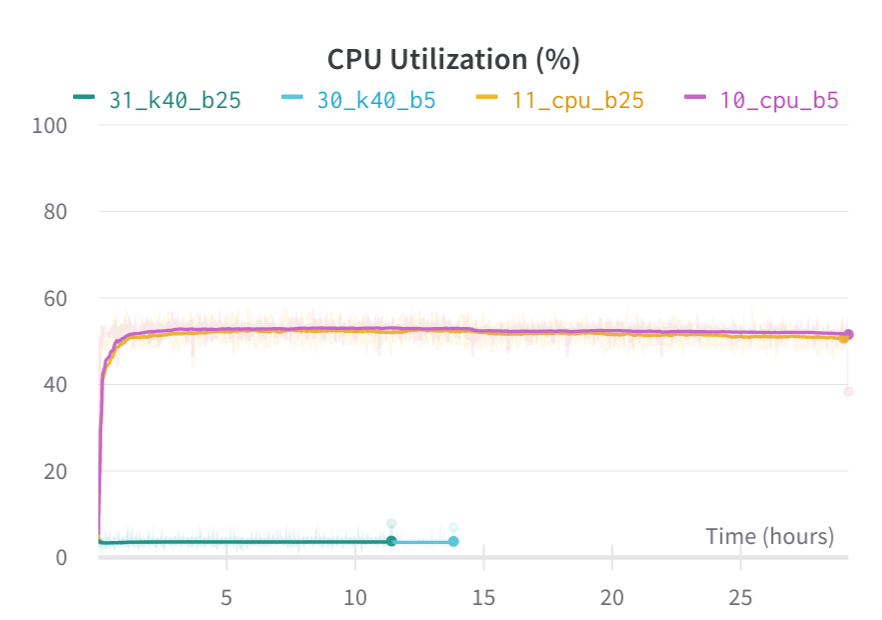}
    \caption{CPU Utilization}
    \label{fig:cpu_utilization}
    \end{minipage}
\end{figure}

The final comparison we have done is between CPU and GPU. With the reference of figure~\ref{fig:24_image_rate}, we can say that image throughput is much better in GPU in comparison to CPU. The figure~\ref{fig:cpu_utilization} shows that the CPU isn't utilizing enough resources and processing a lower number of images per second. The number of parallel instances that the CPU can handle became the bottleneck of the system.  

\begin{figure}[!tbp]
  \centering
  \begin{minipage}[b]{0.32\textwidth}
    \includegraphics[width=\textwidth]{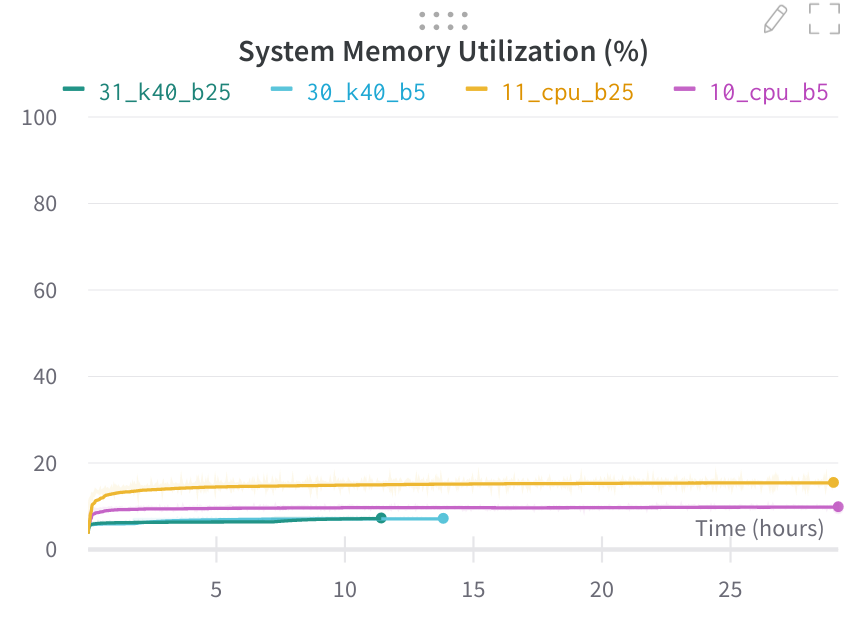}
    \caption{System Memory Utilization (CPU vs K40)}
    \label{fig:24_system_memory}
  \end{minipage}
  \hfill
  \begin{minipage}[b]{0.15\textwidth}
    \includegraphics[width=\textwidth]{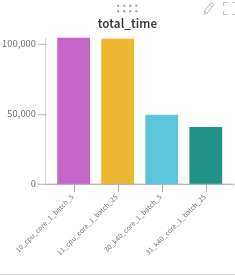}
    \caption{CPU vs GPU Time comparison}
    \label{fig:cpu_vs_gpu_time}
    \end{minipage}
\end{figure}



\section{Conclusion}
Based on the result we observed, we can conclude that memory is the neck of the GPU system when dealing with larger images as an input to a deep neural network model. So, the selection of the GPU should be based on the time available to perform training and the total cost available to spend on the GPU machines. 

In addition, the CPU has limited parallelization capability in comparison to the GPU. As a result, image throughput is better with GPU, and because the CPU doesn't utilize as many resources, it produces fewer images per second. Therefore CPU use doesn't provide an efficient solution when training larger image-based neural networks.

\ifCLASSOPTIONcompsoc
  \section*{Acknowledgments}
\else
  \section*{Acknowledgment}
\fi

The authors would like to thank Professor Dr. Rong Ge for her constant feedback regarding the project. In addition, the authors would like to thank Palmetto Supports for their guidance regarding cluster setup and Palmetto Cluster for providing computers to run this project.



%

\end{document}